\documentclass[pra,twocolumn,showpacs]{revtex4}
\usepackage{epsfig}
\usepackage{graphicx}
\usepackage{dcolumn}
\usepackage{amsmath,amssymb}
\usepackage{pstricks}
\usepackage{color}
\usepackage{footnote}
\usepackage{bm}
\usepackage{bbold}
\usepackage[colorlinks=true,citecolor={blue},urlcolor={blue}, linkcolor=blue]{hyperref}
\begin{document}
\title{Calculation of the magnetic hyperfine structure constant of alkali metals and alkaline earth metal ions using the relativistic coupled-cluster method}
\author{Sudip Sasmal\footnote{sudipsasmal.chem@gmail.com}}
\affiliation{Electronic Structure Theory Group, Physical Chemistry Division, CSIR-National Chemical Laboratory, Pune, 411008, India}
\begin{abstract}
The Z-vector method in the relativistic coupled-cluster framework is used to calculate magnetic hyperfine structure constant
($A_J$) of alkali metals and singly charged alkaline earth metals in their ground state electronic configuration.
The Z-vector results are in very good agreement with the experiment. The $A_J$ values of Li, Na, K, Rb, Cs,
Be$^{+}$, Mg$^{+}$, Ca$^{+}$, and Sr$^{+}$ obtained in the Z-vector method are compared with the extended
coupled-cluster results taken from \newblock{Phys. Rev. A {\bf 91}, 022512 (2015)}. The same basis and cutoff are used for the
comparison purpose. The comparison shows that Z-vector
method with the singles and double approximation can produce more precise wavefunction in the nuclear region than the
ECC method.
\end{abstract}
\maketitle
\section{Introduction}
The interaction of electromagnetic field of electrons with the nuclear moments of nucleus, known as hyperfine structure
interaction, causes small shift and splitting of energy levels \cite{lindgren_book}. Therefore, it is very important for the accurate
description of energy levels of atom, molecule and ion. The precise measurements of the energy levels of alkali
metals play important role in various areas of atomic and nuclear physics as they are extensively used 
in high precession spectroscopy, laser cooling and trapping of atom, ultracold
collision studies, photo-association spectroscopy, Bose-Einstein condensation and more recently, test for
parity and time reversal violation.
Currently, the hyperfine transition of Cs atom [[Xe]6S($^{2}S_{1/2},F=3,m_F=0$) $\leftrightarrow$
[Xe]6S($^{2}S_{1/2},F=4,m_F=0$)] is used as frequency standard which is accurate up to
1 per $10^{15}$ \cite{cs_precession}.
Singly ionized alkaline earth metal ions are insensitive to the perturbation of the environment arising form
collisions and Doppler shift and thus, have been considered for the potential candidates for optical frequency
standard \cite{champenois_2004, barwood_1999, margolis_2004, sherman_2005}.
The $^{2}S_{1/2}$ ground state of these ions are regarded for quantum information processing studies
to encode quantum-bit into hyperfine levels because of their long phase coherence due to their small energy gap
and relatively large spontaneous decay lifetime \cite{ozeri_2007, roos_2004}.
\par
Recently, experiments for parity non-conservation (PNC) become a cutting-edge topic as it can test the accuracy of
fundamental physics and explore ``new physics'' beyond the standard model.
However, the PNC amplitudes, which are very essential to determine the value of PNC constants cannot be measured
experimentally and thus, has to be obtained theoretically. Therefore, it is extremely important to have a reliable
way of determining the accuracy of such theoretical calculations. The PNC amplitudes are very sensitive to the
accuracy of the wavefunction in the near nuclear region \cite{flambaum_1980,flambaum_1984}.
The same is true for HFS constants \cite{tpdas_1987}. Therefore, one can assess
the accuracy of PNC amplitudes by comparing theoretically obtained HFS constants with corresponding
experimental values \cite{kozlov_1997_baf,porsev_1999,AIC}.
\par
Relativistic effects are very important for the precise calculation of the wavefunction in the near nuclear region.
For a single determinant theory, the best way to include relativistic
effect is to solve the four-component Dirac-Hartree-Fock (DHF) equation. However, the DHF method misses the
instantaneous interaction of opposite spin electrons. Coupled-cluster (CC) \cite{cizek_1966, cizek_1967, bartlett_1978}
is the most elegant method to include this dynamic electron correlation.
\par
The coupled-cluster equation can be solved either by variationally or by non-variationally. Although, the
non-variational coupled-cluster, also known as normal coupled cluster (NCC) is the most familiar, the variational
coupled-cluster (VCC)  has several advantages over the NCC. The VCC, being variational has upper boundedness in energy
and satisfies the generalized Hellmann-Feynman (GHF) theorem which simplifies the calculations for higher order properties.
Unitary coupled-cluster (UCC) \cite{kutzelnigg_ucc, pal1983use, pal1984ucc, tanaka_1984, hoffmann_1988, bartlett_ucc},
expectation value coupled-cluster (XCC) \cite{pal_1982, pal1984xcc, bartlett_xcc, ghose_1993}
and extended coupled-cluster (ECC) \cite{arponen_ecc, bishop_ecc} are
the most familiar VCC \cite{szalay_1995} in literature. Recently, ECC has been extended to the relativistic regime to calculate
magnetic HFS constants of atoms and molecules \cite{sasmal_ecc}. ECC uses dual space of right and left vectors in a double linked
form where the left vector is not complex conjugate of the right vector. Although ECC functional is a terminating
series, the natural termination leads to very expensive terms. Thus, for practical purpose, one needs to use some
truncation scheme to avoid computationally expensive terms.
\par
On the other hand, the NCC is nonvariational and thus, does not satisfy the GHF theorem. Therefore, the expectation
value and first order energy derivative yield different results \cite{monkhorst_1977, sekino_1984}.
However, the energy derivative method is superior
than the expectation value method as the property value obtained in energy derivative method can be expressed as
the corresponding expectation value plus some additional terms which make it closer to the full configuration
interaction property value. The NCC energy is not optimized with respect to determinantal coefficients ($C_d$) in the
expansion of many electron wavefunction \cite{monkhorst_1977}. Thus, the derivative of energy with respect to external perturbation requires
the derivative of energy with respect to $C_d$ times the derivative of $C_d$ with respect to external perturbation.
The derivative terms involving $C_d$ can be included in Z-vector \cite{schafer_1984, zvector_1989} method by introducing a perturbation independent
de-excitation operator where the equation for this operator is linear. Thus, for any number of property calculation, one
needs to calculate only one set of
coupled-cluster amplitudes. The advantage of Z-vector method over ECC method is that unlike ECC, the equations for
excitation operators are decoupled from the de-excitation operators. This saves enormous computational cost.
Recently, Z-vector method is extended to the relativistic region for the calculation of ground state properties of
atomic and molecular systems \cite{sasmal_srf}.
\par
In this paper, we have calculated the magnetic HFS constant of alkali metal atoms and singly charged alkaline earth metal cations
using Z-vector technique in the relativistic coupled-cluster framework.
We have compared the Z-vector values with the ECC values calculated
in Ref. \cite{sasmal_ecc} to show that the Z-vector method with single and double approximation
can produce much better wavefunction in the nuclear region of
atomic nucleus than the ECC method with the approximation stated in the paper and thus, is capable of providing
the precise value of the types of property like PNC amplitudes,
which are prominent in the nuclear region.
The paper is organize as follows. A brief introduction and the workable equations for the
Z-vector method are given in Sec. \ref{theory} followed by the matrix elements for the magnetic HFS constant of atomic system
in the same section. The computational details are given in Sec. \ref{comp}. In Sec. \ref{res_dis}, we present our results and
discuss about those before making our final remarks in Sec. \ref{conc}.
\section{Theory}\label{theory}
\subsection{Z-vector method}\label{corr}
The study of hyperfine interaction helps us to understand nuclear structure of an atom and its impact
on the electronic wavefunction in the nuclear and near nuclear region. Therefore, for the accurate
calculation of magnetic HFS constant, which demands very precise wavefunction in the short range of
nucleus, we need to incorporate both relativistic and electron correlation effects. In this work,
the four component Dirac-Hartree-Fock (DHF) method is used to include the effect of relativity where
the electron-electron repulsion term is approximated as Coulomb interaction. The Dirac-Coulomb
Hamiltonian is given by
\begin{eqnarray}
{H_{DC}} &=&\sum_{i} \Big [-c (\vec {\alpha}\cdot \vec {\nabla})_i + (\beta -{\mathbb{1}_4}) c^{2} + V^{nuc}(r_i)+ \nonumber\\
       && \sum_{j>i} \frac{1}{r_{ij}} {\mathbb{1}_4}\Big],
\end{eqnarray}
where, {\bf$\alpha$} and $\beta$ are the usual Dirac matrices, $c$ is the speed of light,
${\mathbb{1}_4}$ is the 4$\times$4 identity matrix and $V^{nuc}(r_i)$ is the nuclear potential
function and the Gaussian charge distribution is used in this work.
The DHF method misses the instantaneous dynamic correlation of opposite spin electrons.
Among various many-body theory, the single reference coupled-cluster (SRCC) is the most elegant
technique to incorporate dynamic correlation. The SRCC wavefunction is given as
\begin{eqnarray}
|\Psi_{cc}\rangle=e^{T}|\Phi_0\rangle ,
\end{eqnarray}
where, $\Phi_0$ is the DHF wavefunction and $T$ is the coupled-cluster excitation operator which is given
by
\begin{eqnarray}
 T=T_1+T_2+\dots +T_N=\sum_n^N T_n ,
\end{eqnarray}
with
\begin{eqnarray}
 T_m= \frac{1}{(m!)^2} \sum_{ij\dots ab \dots} t_{ij \dots}^{ab \dots}{a_a^{\dagger}a_b^{\dagger} \dots a_j a_i}.
\end{eqnarray}
Here, i,j(a,b) are the hole(particle) indices and $t_{ij..}^{ab..}$ are the cluster amplitudes corresponding 
to the cluster operator $T_m$.
In the coupled-cluster single and double (CCSD) approximation, $T=T_1+T_2$. The equations for T$_1$ and T$_2$ are
given as
\begin{eqnarray}
 \langle \Phi_{i}^{a} | (H_Ne^T)_c | \Phi_0 \rangle = 0 , \nonumber\\
  \langle \Phi_{ij}^{ab} | (H_Ne^T)_c | \Phi_0 \rangle = 0 ,
 \label{cc_amplitudes}
\end{eqnarray}
where, H$_N$ is the normal ordered DC Hamiltonian and subscript $c$ means only the connected terms exist in the
contraction between H$_N$ and T. Size-extensivity is ensured by this connectedness.
The coupled-cluster correlation energy can be obtained as
\begin{eqnarray}
 E^{corr} = \langle \Phi_0 | (H_Ne^T)_c | \Phi_0 \rangle .
\label{cc_energy}
\end{eqnarray}
\par
However, the SRCC energy is not optimized with respect to the determinantal coefficients and the molecular
orbital coefficients in the expansion of the many electron correlated wavefunction \cite{monkhorst_1977}.
Therefore,
the calculation of SRCC energy derivative with respect to external perturbation requires to include
these derivative terms. The equation for these terms are linear but in general,
perturbation dependent. However, in Z-vector method, the derivative terms containing the determinantal coefficients
can be
incorporated by the introduction of a perturbation independent operator $\Lambda$ \cite{zvector_1989}. Thus, in the Z-vector
method, any number of property calculations can be done by solving only one set of $T$ and $\Lambda$
amplitudes.
$\Lambda$ is a deexcitation operator and the second quantized form is given by
\begin{eqnarray}
 \Lambda=\Lambda_1+\Lambda_2+...+\Lambda_N=\sum_n^N \Lambda_n ,
\end{eqnarray}
where,
\begin{eqnarray}
 \Lambda_m= \frac{1}{(m!)^2} \sum_{ij..ab..} \lambda_{ab..}^{ij..}{a_i^{\dagger}a_j^{\dagger} .. ..a_b a_a}.
\end{eqnarray}
Here $\lambda_{ab..}^{ij..}$ are the cluster amplitudes corresponding to the cluster operator $\Lambda_m$.
In CCSD approximation, $\Lambda=\Lambda_1+\Lambda_2$. The explicit equations for the amplitudes of $\Lambda_1$
and $\Lambda_2$ operators are given by
\begin{eqnarray}
\langle \Phi_0 |[\Lambda (H_Ne^T)_c]_c | \Phi_{i}^{a} \rangle + \langle \Phi_0 | (H_Ne^T)_c | \Phi_{i}^{a} \rangle = 0,
\label{lambda_1}
\end{eqnarray}
\begin{eqnarray}
\langle \Phi_0 |[\Lambda (H_Ne^T)_c]_c | \Phi_{ij}^{ab} \rangle + \langle \Phi_0 | (H_Ne^T)_c | \Phi_{ij}^{ab} \rangle \nonumber \\
 + \langle \Phi_0 | (H_Ne^T)_c | \Phi_{i}^{a} \rangle \langle \Phi_{i}^{a} | \Lambda | \Phi_{ij}^{ab} \rangle = 0.
\label{lambda_2}
\end{eqnarray}
The energy derivative is given by
\begin{eqnarray}
 \Delta E' = \langle \Phi_0 | (O_Ne^T)_c | \Phi_0 \rangle + \langle \Phi_0 | [\Lambda (O_Ne^T)_c]_c | \Phi_0 \rangle .
\end{eqnarray}
Here, $O_N$ is the derivative of normal ordered perturbed Hamiltonian with respect to external field of perturbation.
It is clear from the above formulation that the derivative terms containing only the determinantal coefficients
are included here, i.e., the orbital relaxation terms that are required to make energy functional stationary
with respect to molecular orbital coefficients are not considered here.
It is worth to mention that
recently, Saue and coworkers \cite{shee_2016} have implemented the orbital-unrelaxed analytical method in the
four-component relativistic SRCC framework based on the Lagrangian multiplier method of Helgaker and coworkers \cite{koch_1990}
which is similar to the Z-vector method for the ground state first order properties.
\subsection{Magnetic hyperfine structure constant}\label{prop}
The magnetic HFS interaction arises due to the coupling of nuclear magnetic moment with the angular momentum
of electrons and thus, can be treated as a one-body interaction from the electronic structure point of view.
The magnetic vector potential due to a nucleus is given by
\begin{equation}
\vec{A}=\frac{\vec{\mu}_k \times \vec{r}}{r^3},
\end{equation}
where, $\vec{\mu}_k$ is the magnetic moment of nucleus $K$.
In Dirac theory, the HFS interaction Hamiltonian due to $\vec{A}$ can be given as
\begin{equation}
H_\text{hfs}= \sum_i^n \vec{\alpha}_i \cdot \vec{A_i}, 
\end{equation}
where, $\alpha_i$ denotes the Dirac $\alpha$ matrices for the i$^{th}$ electron and $n$ is the total no of electrons.
The magnetic hyperfine constant of the $J^{th}$ electronic state of an atom can be given as
\begin{eqnarray}
 A_J &=& \frac{1}{IJ} \langle \Psi_J | H_\text{hfs} | \Psi_J \rangle \nonumber \\
     &=& \frac{\vec{\mu}_k}{IJ} \cdot \langle \Psi_J | \sum_i^n \left( 
       \frac{\vec{\alpha}_i \times \vec{r}_i}{r_i^3} \right) | \Psi_J \rangle,
 \label{hfs_atom}
\end{eqnarray}
where, $I$ is the nuclear spin quantum number and $\Psi_J$ is the wavefunction of the $J^{th}$ electronic state.
\section{Computational details}\label{comp}
\begin{table}[ht]
\caption{Basis and cutoff used for the atomic calculation.}
\begin{ruledtabular}
\newcommand{\mc}[1]{\multicolumn{#1}}
\begin{center}
\begin{tabular}{lcr}
Atom & Basis & Virtual cutoff (a.u.) \\
\hline
Li & aug-cc-pCVQZ & × \\
Na & aug-cc-pCVQZ & × \\
K & dyall.cv4z & 500 \\
Rb & dyall.cv3z & 500 \\
Cs & dyall.cv4z & 40 \\
Fr & dyall.cv3z & 50 \\
Be$^{+}$ & aug-cc-pCVQZ & × \\
Mg$^{+}$ & aug-cc-pCVQZ & × \\
Ca$^{+}$ & dyall.cv4z & 500 \\
Sr$^{+}$ & dyall.cv3z & 100 \\
Ba$^{+}$ & dyall.cv4z & 40 \\
Ra$^{+}$ & dyall.cv3z & 50 
\end{tabular}
\end{center}
\label{basis_atom}
\end{ruledtabular}
\end{table}
The DIRAC10 program package \cite{dirac10} is used to solve the DHF equation and to construct the one-electron
and two-electron matrix elements. The magnetic HFS integrals are extracted from a locally modified version of
DIRAC10. Gaussian charge distribution is considered for the finite size of the nucleus where the the nuclear
parameters are taken from Ref. \cite{visscher_1997}. Restricted kinetic balance \cite{dyall_2007}
condition is used to link small
and large component basis function. No virtual pair approximation (NVPA) is used to solve DHF equation.
This means that the negative energy solutions are removed by using projection operator and only positive
energy solutions are included in the correlation calculations. However, how to go beyond the no-pair approximation
by accounting for correlation contributions of negative energy states has been discussed in depth in Ref. \cite{liu_2013, liu_2014, liu_2016}.
In our calculation, we have used aug-cc-pCVQZ basis \cite{ccpcvxz_h_b-ne, aug_ccpcvxz_NaMg} for Li, Na, Be and Mg atoms
and dyall.cv3z basis \cite{dyall_s} for Rb, Sr, Fr and Ra
atoms and dyall.cv4z \cite{dyall_s} basis for K, Ca, Cs and Ba atoms. All electrons are considered for the correlation calculation
of all systems. The cutoff used for the virtual orbitals are compiled in Table \ref{basis_atom}.
\section{Results and discussion}\label{res_dis}
\begin{table}[ht]
\scriptsize
\caption{Magnetic hyperfine coupling constant (in MHz) of ground state of atoms.}
\begin{ruledtabular}
\newcommand{\mc}[1]{\multicolumn{#1}}
\begin{center}
\begin{tabular}{lcccrcr}
Atom             & SCF & ECC & Z-vector & Expt.& \mc{2}{c}{$\delta$\%}\\
\cline{6-7}
 & & \cite{sasmal_ecc} & & & ECC & Z-vector\\
\hline
$^{6}$Li         & 107.2 & 149.3 & 148.3 & 152.1 \cite{beckmann_1974} & 1.9 & 2.6 \\
$^{7}$Li         & 283.2 & 394.3 & 391.6 & 401.7 \cite{beckmann_1974} & 1.9 & 2.6 \\
$^{23}$Na        & 630.6 & 861.8 & 861.4 & 885.8 \cite{beckmann_1974} & 2.8 & 2.8 \\
$^{39}$K         & 151.0 & 223.5 & 226.6 & 230.8 \cite{beckmann_1974} & 3.3 & 1.9 \\
$^{40}$K         & -187.7 & -277.9 & -281.8 & -285.7 \cite{eisinger_1952} & 2.8 & 1.4 \\
$^{41}$K         & 82.9 & 122.7 & 124.4 & 127.0 \cite{beckmann_1974} & 3.5 & 2.1 \\
$^{85}$Rb        & 666.9 & 972.5 & 986.5 & 1011.9 \cite{vanier_1974} & 4.1 & 2.6 \\
$^{87}$Rb        & 2260.1 & 3295.7 & 3343.3 & 3417.3 \cite{essen_1961} & 3.7 & 2.2 \\
$^{133}$Cs       & 1495.5 & 2179.1 & 2218.4 & 2298.1 \cite{arimondo_1977} & 5.5 & 3.6 \\
$^{223}$Fr       & 5518.0 & × & 7537.4 & 7654(2)\cite{sansonetti_2007}  & × & 1.5 \\
$^{9}$Be$^{+}$   & -498.8 & -614.6 & -612.9 & -625.0 \cite{wineland_1983} & 1.7 & 2.0 \\
$^{25}$Mg$^{+}$  & -466.7 & -581.6 & -584.8 & -596.2 \cite{itano_1981} & 2.5 & 1.9 \\
$^{43}$Ca$^{+}$  & -606.2 & -794.9 & -801.5 & -806.4 \cite{arbes_1994} & 1.4 & 0.6 \\
$^{87}$Sr$^{+}$  & -761.0 & -969.9 & -977.9 & -1000.5(1.0) \cite{buchinger_1990} & 3.2 & 2.3 \\
$^{135}$Ba$^{+}$ & 2737.4 & × & 3513.3 & 3591.7 \cite{trapp_2000} & × & 2.2 \\
$^{137}$Ba$^{+}$ & 3062.1 & × & 3930.2 & 4018.9 \cite{trapp_2000} & × & 2.3 \\
$^{223}$Ra$^{+}$ & 2842.8 & × & 3446.3 & 3404(2) \cite{wendt_1987,neu_1989} & × & 1.2 \\
\end{tabular}
\end{center}
\label{atom_hfs}
\end{ruledtabular}
\end{table}
In Table \ref{atom_hfs}, we present the magnetic HFS constant of alkali metal atoms and mono-positive
alkaline earth metal ions in their ground state ($^{2}S$) electronic configuration using the Z-vector technique
in the relativistic coupled-cluster framework. We have compiled the experimental values for these systems
in the same table and the relative deviations of Z-vector results from the experimental values are presented
as $\delta$\%.
The results for different isotopes are calculated by using their corresponding nuclear magnetic moment values
although the nuclear parameters in the nuclear model are taken as same for each isotope which is default the most
stable isotopes in DIRAC10 \cite{dirac10}.
Our calculated Z-vector results are in very good agreement with the experimental values.
From the table, it is clear that the deviations of Z-vector results from the experiments are well within 3\% except
for the $^{133}$Cs atom, where it is 3.6\%. The Z-vector results are quite impressive, especially for the heavy atoms.
\begin{figure}[ht]
\centering
\begin{center}
\includegraphics[scale=.1, height=5cm]{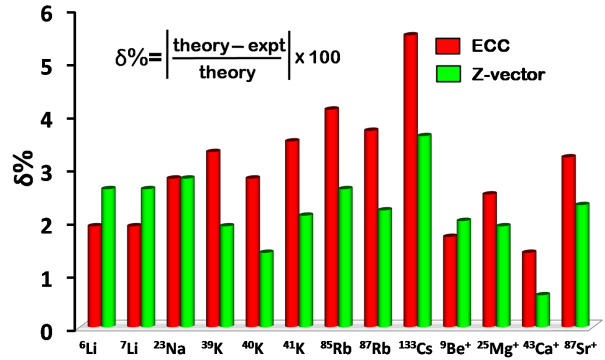}
\caption {Comparison of relative deviations between Z-vector and ECC values of magnetic HFS constant of atoms.}
\label{zvec_atom_comp}
\end{center}  
\end{figure}
The ECC values of magnetic HFS constant are taken from Ref. \cite{sasmal_ecc} and the deviations from the experiment
are presented in the table. We have used same basis and cutoff for those systems for comparison purpose.
The deviations of ECC and Z-vector values from the experimental values are presented in Fig. \ref{zvec_atom_comp}.
From the figure, it is clear that Z-vector results are far better than the ECC results except for two small systems
like Li and Be$^{+}$.
As the magnetic HFS constant is very sensitive to the near nuclear wavefunction,
the above results show that the Z-vector method can produce far better wavefunction in
the nuclear region than the ECC method and the results are quite impressive for the heavy atoms.
Although, ECC is a truncated series, in CCSD model, the natural truncation leads to very expensive terms.
In Ref. \cite{sasmal_ecc}, the truncation scheme proposed by Vaval {\it et al} are used to avoid the expensive
terms in the ECC functional where the right exponential is full within the CCSD approximation and the higher-order
double-linked terms within the CCSD approximation are taken in the left exponent. This approximation introduces
an additional error which may be the reason for the poor performance of ECC compared to Z-vector method.
\begin{table}[h]
\caption{ Comparison of full CI and Z-vector magnetic HFS values (in MHz) of $^{7}$Li}
\begin{ruledtabular}
\begin{tabular}{lcr}
Basis & Full CI \cite{sasmal_ecc} & Z-vector\\
\hline
aug-cc-pCVDZ &\, 384.1  &\, 383.9 \\
aug-cc-pCVTZ &\, 402.0 &\, 401.3 \\
aug-cc-pCVQZ$^{a}$ &\, 386.0 &\, 385.2
\end{tabular}
\end{ruledtabular}
\label{li}
$^{a}$ Considering 3 electrons and 189 virtual orbitals
\end{table}
\begin{table}[h]
\caption{Comparison of full CI and Z-vector magnetic HFS values (in MHz) of $^{9}$Be$^{+}$}
\begin{ruledtabular}
\begin{tabular}{lcr}
Basis & Full CI \cite{sasmal_ecc} & Z-vector\\
\hline
aug-cc-pCVDZ &\, -586.6  &\, -586.5 \\
aug-cc-pCVTZ &\, -615.7 &\, -615.4 \\
aug-cc-pCVQZ$^{a}$ &\, -613.0 &\, -612.7
\end{tabular}
\end{ruledtabular}
\label{be}
$^{a}$ Considering 3 electrons and 183 virtual orbitals
\end{table}
\par
The HFS constant predominantly depends on the spin density of the valence electron in the nuclear
region and thus is not very sensitive to the
retardation and magnetic effects described by the Breit interaction \cite{quiney_1998_tlf,lindroth_1989}.
It can be seen from the previous calculations that the higher order relativistic effects on these types of properties
generally lie $\sim$ 0.5-1\% \cite{blundell_1990,kozlov_2001,dzuba_2002}.
It is worth to mention here that although we have correlated all electrons in our Z-vector calculations, the results are not
completely free from the uncertainty associated with core correlation as the cvNz (N=3,4) basis set misses some important
core correlating functions. The 1s-3d electrons also need much higher virtual energy orbitals for proper correlation functions
as shown in Ref. \cite{skripnikov_2016,skripnikov_2017}.
A series of calculations are done to estimate the uncertainty associated with the Z-vector values of the magnetic HFS constant of these systems.
The comparison between full configuration interaction (FCI) and Z-vector magnetic HFS constant values of $^{7}$Li and $^{9}$Be$^{+}$  has been
made and is
presented in Tables \ref{li} and \ref{be}, respectively. By comparing Z-vector values with FCI values and considering all
other sources of error like higher
order relativistic effects, missing correlation effects etc, it can be assumed that the overall uncertainty in our final results is less than 5\%.
\section{Conclusion}\label{conc}
We have calculated the magnetic HFS constant of alkali metal atoms (Li, Na, K, Rb, Cs and Fr) and mono-positive
alkaline earth metal ions (Be$^{+}$, Mg$^{+}$, Ca$^{+}$, Sr$^{+}$, Ba$^{+}$ and Ra$^{+}$) using the
Z-vector technique in the relativistic coupled-cluster framework.
We have compared the Z-vector values and
the ECC values taken from Ref. \cite{sasmal_ecc} with experiment and the comparison shows that the Z-vector
method with single and double approximation can produce much accurate wavefunction in the nuclear region than the
ECC method with the given approximation.
A possible explanation
for the poor performance of the ECC method is also given.
\section*{Acknowledgement}
Author thanks Prof. Sourav Pal, Dr. Malaya K. Nayak, Dr. Nayana Vaval, Dr. Himadri Pathak for their valuable comments
and insights in this work. Author acknowledges Dr. Malaya K. Nayak for providing the hyperfine integrals.
Author acknowledges the resources of the Center of Excellence in Scientific Computing at CSIR-NCL.
S.S. acknowledges the CSIR for the SPM fellowship. 

\begin{thebibliography}{10}

\bibitem{lindgren_book}
I.~Lindgren and J.~Morrison,
\newblock {\em Atomic Many-Body Theory} (Springer-Verlag, New York, 1985).

\bibitem{cs_precession}
http://tf.nist.gov/cesium/atomichistory.htm.

\bibitem{champenois_2004}
C.~Champenois {\em et~al.},
\newblock Physics Letters A {\bf 331}, 298 (2004).

\bibitem{barwood_1999}
G.~P.~Barwood, G.~Huang, H.~A.~Klein, P.~Gill, and R.~B.~M.~Clarke,
\newblock Physical Review A {\bf 59}, R3178 (1999).

\bibitem{margolis_2004}
H.~Margolis {\em et~al.},
\newblock Science {\bf 306}, 1355 (2004).

\bibitem{sherman_2005}
J.~A. Sherman, W.~Trimble, S.~Metz, W.~Nagourney, and N.~Fortson,
\newblock Progress on indium and barium single ion optical frequency standards,
\newblock in {\em LEOS Summer Topical Meetings, 2005 Digest of the}, pp.
  99--100, IEEE, 2005.

\bibitem{ozeri_2007}
R.~Ozeri, W.~M.~Itano, R.~B.~Blakestad, J.~Britton, J.~Chiaverini, J.~D.~Jost, C.~Langer, D.~Leibfried, R.~Reichle, S.~Seidelin,
J.~H.~Wesenberg, and D.~J.~Wineland,
\newblock Physical Review A {\bf 75}, 042329 (2007).

\bibitem{roos_2004}
C.~F. Roos {\em et~al.},
\newblock Science {\bf 304}, 1478 (2004).

\bibitem{flambaum_1980}
V.~Flambaum and I.~Khriplovich,
\newblock Sov. Phys.-JETP (Engl. Transl.);(United States) {\bf 52} (1980).

\bibitem{flambaum_1984}
V.~Flambaum, I.~Khriplovich, and O.~Sushkov,
\newblock Physics Letters B {\bf 146}, 367 (1984).

\bibitem{tpdas_1987}
T.~Das,
\newblock Hyperfine Interactions {\bf 34}, 149 (1987).

\bibitem{kozlov_1997_baf}
M.~G.~Kozlov, A.~V.~Titov, N.~S.~Mosyagin, and P.~V.~Souchko,
\newblock Physical Review A {\bf 56}, R3326 (1997).

\bibitem{porsev_1999}
S.~Porsev, Y.~G. Rakhlina, and M.~Kozlov,
\newblock Journal of Physics B: Atomic, Molecular and Optical Physics {\bf 32},
  1113 (1999).

\bibitem{AIC}
A.~V. Titov, Y.~V. Lomachuk, and L.~V. Skripnikov,
\newblock Phys. Rev. A {\bf 90}, 052522 (2014).

\bibitem{cizek_1966}
J.~{\v{C}}{\'\i}{\v{z}}ek,
\newblock Journal of Chemical Physics {\bf 45}, 4256 (1966).

\bibitem{cizek_1967}
J.~Cizek,
\newblock {\em Advances in Chemical Physics: Correlation Effects in Atoms and
  Molecules} (Wiley, Hoboken, NJ, 1967).

\bibitem{bartlett_1978}
R.~J. Bartlett and G.~D. Purvis,
\newblock International Journal of Quantum Chemistry {\bf 14}, 561 (1978).

\bibitem{kutzelnigg_ucc}
W.~Kutzelnigg,
\newblock {\em Methods of Electronic Structure Theory} (Plenum, New York,
  1977).

\bibitem{pal1983use}
S.~Pal, M.~D. Prasad, and D.~Mukherjee,
\newblock Theoretica chimica acta {\bf 62}, 523 (1983).

\bibitem{pal1984ucc}
S.~Pal,
\newblock Theoretica chimica acta {\bf 66}, 207 (1984).

\bibitem{tanaka_1984}
K.~Tanaka and H.~Terashima,
\newblock Chemical physics letters {\bf 106}, 558 (1984).

\bibitem{hoffmann_1988}
M.~R. Hoffmann and J.~Simons,
\newblock Journal of Chemical Physics {\bf 88}, 993 (1988).

\bibitem{bartlett_ucc}
R.~J. Bartlett, S.~A. Kucharski, and J.~Noga,
\newblock Chemical physics letters {\bf 155}, 133 (1989).

\bibitem{pal_1982}
S.~Pal, M.~D. Prasad, and D.~Mukherjee,
\newblock Pramana {\bf 18}, 261 (1982).

\bibitem{pal1984xcc}
S.~Pal,
\newblock Theoretica chimica acta {\bf 66}, 151 (1984).

\bibitem{bartlett_xcc}
R.~J. Bartlett and J.~Noga,
\newblock Chemical physics letters {\bf 150}, 29 (1988).

\bibitem{ghose_1993}
K.~B. Ghose, P.~G. Nair, and S.~Pal,
\newblock Chemical physics letters {\bf 211}, 15 (1993).

\bibitem{arponen_ecc}
J.~Arponen,
\newblock Annals of Physics {\bf 151}, 311 (1983).

\bibitem{bishop_ecc}
R.~Bishop, J.~Arponen, and P.~Pajanne,
\newblock {\em Aspects of Many-body Effects in Molecules and Extended Systems,
  Lecture Notes in Chemistry Vol. 50} (Springer-Verlag, Berlin, 1989).

\bibitem{szalay_1995}
P.~G. Szalay, M.~Nooijen, and R.~J. Bartlett,
\newblock Journal of Chemical Physics {\bf 103}, 281 (1995).

\bibitem{sasmal_ecc}
S.~Sasmal, H.~Pathak, M.~K. Nayak, N.~Vaval, and S.~Pal,
\newblock Phys. Rev. A {\bf 91}, 022512 (2015).

\bibitem{monkhorst_1977}
H.~J. Monkhorst,
\newblock Int. J. Quantum Chem. {\bf 12}, 421 (1977).

\bibitem{sekino_1984}
H.~Sekino and R.~J. Bartlett,
\newblock International Journal of Quantum Chemistry {\bf 26}, 255 (1984).

\bibitem{schafer_1984}
N.~C. Handy and H.~F. Schaefer,
\newblock Journal of Chemical Physics {\bf 81}, 5031 (1984).

\bibitem{zvector_1989}
E.~A. Salter, G.~W. Trucks, and R.~J. Bartlett,
\newblock Journal of Chemical Physics {\bf 90}, 1752 (1989).

\bibitem{sasmal_srf}
S.~Sasmal, H.~Pathak, M.~K. Nayak, N.~Vaval, and S.~Pal,
\newblock Phys. Rev. A {\bf 91}, 030503 (2015).

\bibitem{shee_2016}
A.~Shee, L.~Visscher, and T.~Saue,
\newblock The Journal of Chemical Physics {\bf 145}, 184107 (2016).

\bibitem{koch_1990}
H.~Koch {\em et~al.},
\newblock Journal of Chemical Physics {\bf 92}, 4924 (1990).

\bibitem{dirac10}
{DIRAC}, a relativistic ab initio electronic structure program, Release
  {DIRAC10} (2010), written by T.~Saue, L.~Visscher and H.~J.~{\relax
  Aa}.~Jensen, with contributions from R.~Bast, K.~G.~Dyall, U.~Ekstr{\"o}m,
  E.~Eliav, T.~Enevoldsen, T.~Fleig, A.~S.~P.~Gomes, J.~Henriksson,
  M.~Ilia{\v{s}}, Ch.~R.~Jacob, S.~Knecht, H.~S.~Nataraj, P.~Norman, J.~Olsen,
  M.~Pernpointner, K.~Ruud, B.~Schimmelpfennig, J.~Sikkema, A.~Thorvaldsen,
  J.~Thyssen, S.~Villaume, and S.~Yamamoto (see {http://www.diracprogram.org}).

\bibitem{visscher_1997}
L.~Visscher and K.~Dyall,
\newblock Atomic Data and Nuclear Data Tables {\bf 67}, 207  (1997).

\bibitem{dyall_2007}
K.~Faegri~Jr and K.~G. Dyall,
\newblock {\em Introduction to relativistic quantum chemistry} (Oxford
  University Press, USA, 2007).

\bibitem{liu_2013}
W.~Liu and I.~Lindgren,
\newblock Journal of Chemical Physics {\bf 139}, 014108 (2013).

\bibitem{liu_2014}
W.~Liu,
\newblock Physics Reports {\bf 537}, 59 (2014).

\bibitem{liu_2016}
W.~Liu,
\newblock National Science Review {\bf 3}, 204 (2016).

\bibitem{ccpcvxz_h_b-ne}
T.~H. Dunning,
\newblock Journal of Chemical Physics {\bf 90}, 1007 (1989).

\bibitem{aug_ccpcvxz_NaMg}
D. E. Woon and T. H. Dunning Jr. (unpublished).

\bibitem{dyall_s}
K.~G. Dyall,
\newblock The Journal of Physical Chemistry A {\bf 113}, 12638 (2009).

\bibitem{beckmann_1974}
A.~Beckmann, K.~B{\"o}klen, and D.~Elke,
\newblock Zeitschrift f{\"u}r Physik {\bf 270}, 173 (1974).

\bibitem{eisinger_1952}
J.~Eisinger, B.~Bederson, and B.~Feld,
\newblock Physical Review {\bf 86}, 73 (1952).

\bibitem{vanier_1974}
J.~Vanier, J.-F. Simard, and J.-S. Boulanger,
\newblock Physical Review A {\bf 9}, 1031 (1974).

\bibitem{essen_1961}
L.~Essen, E.~Hope, and D.~Sutcliffe,
\newblock Nature {\bf 189}, 298 (1961).

\bibitem{arimondo_1977}
E.~Arimondo, M.~Inguscio, and P.~Violino,
\newblock Reviews of Modern Physics {\bf 49}, 31 (1977).

\bibitem{sansonetti_2007}
J.~E. Sansonetti,
\newblock Journal of physical and chemical reference data {\bf 36}, 497 (2007).

\bibitem{wineland_1983}
D.~J. Wineland, J.~J. Bollinger, and W.~M. Itano,
\newblock Physical Review Letters {\bf 50}, 628 (1983).

\bibitem{itano_1981}
W.~M. Itano and D.~J. Wineland,
\newblock Physical Review A {\bf 24}, 1364 (1981).

\bibitem{arbes_1994}
F.~Arbes, M.~Benzing, T.~Gudjons, F.~Kurth, and G.~Werth,
\newblock Zeitschrift f{\"u}r Physik D Atoms, Molecules and Clusters {\bf 31},
  27 (1994).

\bibitem{buchinger_1990}
F. Buchinger, E. B. Ramsay, E. Arnold, W. Neu, R. Neugart, K. Wendt, R. E. Silverans, P. Lievens, L. Vermeeren,
D. Berdichevsky, R. Fleming, D. W. L. Sprung, and G. Ulm,
\newblock Physical Review C {\bf 41}, 2883 (1990).

\bibitem{trapp_2000}
S.~Trapp {\em et~al.},
\newblock Hyperfine Interactions {\bf 127}, 57 (2000).

\bibitem{wendt_1987}
K.~Wendt {\em et~al.},
\newblock Zeitschrift f{\"u}r Physik D Atoms, Molecules and Clusters {\bf 4},
  227 (1987).

\bibitem{neu_1989}
W.~Neu {\em et~al.},
\newblock Zeitschrift f{\"u}r Physik D Atoms, Molecules and Clusters {\bf 11},
  105 (1989).

\bibitem{quiney_1998_tlf}
H.~M. Quiney, J.~K. Laerdahl, K.~F{\ae}gri~Jr, and T.~Saue,
\newblock Physical Review A {\bf 57}, 920 (1998).

\bibitem{lindroth_1989}
E.~Lindroth, B.~Lynn, and P.~Sandars,
\newblock Journal of Physics B: Atomic, Molecular and Optical Physics {\bf 22},
  559 (1989).

\bibitem{blundell_1990}
S.~A.~Blundell, W.~R.~Johnson, and J.~Sapirstein,
\newblock Physical review letters {\bf 65}, 1411 (1990).

\bibitem{kozlov_2001}
M.~G.~Kozlov, S.~G.~Porsev, and I.~I.~Tupitsyn,
\newblock Physical review letters {\bf 86}, 3260 (2001).

\bibitem{dzuba_2002}
V.~A.~Dzuba, V.~V.~Flambaum, and J.~S.~M.~Ginges,
\newblock Physical Review D {\bf 66}, 076013 (2002).

\bibitem{skripnikov_2016}
L.~V.~Skripnikov,
\newblock The Journal of Chemical Physics {\bf 145}, 214301 (2016).

\bibitem{skripnikov_2017}
L.~V.~Skripnikov, D.~E.~Maison, and N.~S.~Mosyagin,
\newblock Physical Review A {\bf 95}, 022507 (2017).

\end{thebibliography}

\end{document}